\documentclass[runningheads]{llncs}
\renewcommand{\figurename}{Table}
\usepackage{graphicx}
\usepackage{verbatim}
\usepackage{amssymb}
\usepackage{booktabs}
\usepackage{longtable}
\usepackage{float}
\restylefloat{table}
\usepackage{xcolor}
\definecolor{Kim}{rgb}{1,0,0}
\definecolor{Sejuti}{rgb}{0.5,1,0}
\definecolor{Irvin}{rgb}{0,0.5,1}
\usepackage[margin=1.13in]{geometry}
\begin{document}
\title{IoT Platforms for 5G Network and Practical Considerations: A Survey}
%
%


\author{Sejuti Banik\inst{1} \and
Irvin Steve Cardenas\inst{1} \and
Jong-Hoon Kim\inst{1}}

\authorrunning{S. Banik et al.}
\institute{Advanced Telerobotics Research Laboratory\\
Kent State University, Kent OH 44242, USA\\ \email{\{sbanik,icardena,jkim72\}@kent.edu}\\
\url{http://www.atr.cs.kent.edu} \}}
\maketitle              
%
\begin{abstract}
The fifth generation (5G) mobile network will enable the Internet of Things (IoT) to take a large leap into the age of future computing. As a result of extended connectivity, high speed, reduced latency services being provided by 5G, IoT has experienced and will continue to undergo a remarkable transition in every field of daily life. Furthermore, fog computing will revolutionize the IoT platforms by decentralizing the operations by the cloud and ensuring sustainability with big data, mobility and reduced processing lag. 5G is ubiquitous, reliable, scalable and economic in nature. The features will not only globalize IoT in a broader spectrum, but also make common people interact smartly and efficiently with the environment in real time. In this study, a combined survey is presented on different IoT applications coupled with cloud platforms. Moreover, the capabilities of IoT in the influence of 5G are explored as well as how the IoT platform and services will adopt through 5G are envisaged. Additionally, some open issues triggered by 5G have been introduced to harness the maximum benefit out of this network. Finally, a platform is proposed to implement in the telepresence project based on the investigation and findings.
\end{abstract}
\section{Introduction}
The Internet of Things (IoT) paradigm will cross new boundaries in the coming decade due to the continous advancement of technology. In simple words, IoT can be defined as the extension of Internet connection to everyday physical devices enabling smart interactions between other devices and people. The activities and interactions of such devices, or things, e.g. sensors and other hardware in the IoT network can be monitored, and controlled remotely and ubiquitously through end devices such as smart phones or personal computers. Progress in IoT architecture(s) and infrastructure(s) has turned the concepts of smart homes, smart medical care, smart transportation and vehicles-to-everything (V2X) communication, smart environment monitoring, smart industries, and smart agriculture into tangible implementation. 


Contemporary IoT platforms provide services that enable the interaction between hardware devices, software and web services that are part of an IoT system. In specific, these platforms provide the middleware necessary to  successfully connect hardware, negotiate between the communication protocols of different hardware and software, access the cloud for storage, perform analysis and processing on IoT data, and also provide services that are aimed at providing privacy and security of communication between devices. Some platforms, even provide device-specific support and software (e.g. Arm MBED, RTOS for MCUs) \cite{ref_url1}. From a hardware perspective, electronic chips like QCA4020 SoC \cite{ref_url2}, by Qualcomm Technology, are interoperable with devices that use different radio technologies; as well as integrate with other existing software platforms, IoT ecosystems and cloud services. According to IoT Analytics \cite{ref_url3}, there are over 450 IoT platforms catering to the changing demands of personal, and commercial usage. 

But as the number of users rises, the confined range of current 4G towers incur bandwidth deficiency, and their accommodation capacity is exceeded. Furthermore, the latency of communication rises to process the huge amount of data generated each second by the enormous pool of devices. The 5th generation network helps to eliminate these limitations. High speed, capacity for larger number of devices without bandwidth constraint, and reduced latency are only few of the features to name. 
The overwhelming number of IoT platforms often leads to confusion over which platform suits the intended purpose of the users. Additionally, different platforms offer different services which are essential in specific use-cases. Hence, this paper attempts to provide a survey on the unique feature set and limitations of some of the most popular platforms and how they will conform to 5G. Additionally, a platform enabled by 5G is proposed for future application in telepresence robots.

Lastly, a brief overview of the privacy issues, their theoretical solutions and the apparent disadvantages of 5G have been discussed to draw the conclusion.

\section{The Paradigm and Vision}
An Internet of Things represents the concept of bridging the physical world with the virtual world - interconnecting everyday physical objects and allowing them to access Internet services \cite{2010_mattern_internet_of_computers_to_iot}. It is a concept that was initially put forth by Mark Weiser in the early 1990's \cite{1991_weiser_the_computer}. Since then, we have seen the advancement of smart devices, network technologies, and a growing push from industry to turn the concept of IoT into a reality. But, although close to three decades have passed since this concept was formally discussed, we are still faced with technical and social questions over the Internet of Things. One prominent technical question relates to the constraints of achieving scalable and secure communication between such devices - through  existing infrastructures and communication protocols. Nonetheless, this has not stopped the growth in the number of commercially-available smart devices such as thermostats, light bulbs, nor deter the drive towards the concepts of smart buildings and smart cities \cite{2015_iot_framework_smart_energy_in_buildings,2015_novel_wireless_sensor_network_urban_transport,2016_iot_connected_nav_sys_urban_bus_riders,2017_iot_considerations_reqs_archs_smart_buildings_energy}. 

In parallel, we have seen the rapid development and adoption of new mobile technology which has brought the need for telecommunication networks to support the ever growing number of mobile devices and their continuous demand for data at faster speeds. It is a problem of scalability and throughput which cannot be adequately handled by existing 4G/LTE networks. 5G networks come as a solution to such need. In all, made possible through a combination of new technologies. Five of the most prominent technologies are: (1) Millimeter Waves  \cite{2015_millimeter_wave_comm,2018_5G_milli_wave_propagation_models_and_perf_eval}, (2) Small Cell Networks \cite{2014_small_cells_mimo,2016_5g_small_cell_nets_fronthauling}, multi-antenna technologies such as (3) Multiple-Input Multiple-Output (MIMO) \cite{2014_MIMO_beamforming} and (4) Beamforming \cite{2014_MIMO_beamforming}, and (5) Full Duplex \cite{2015_full_duplex_techniques,2018_5g_enabling_tech_benefits_feasibility_limitations_full_duplex_mMIMO}. 5G is set to be the foundation for many other emerging technologies, from virtual reality, to autonomous driving, mission critical application.

In all, the introduction of 5G network technology and the convergence of the IoT paradigm with new network paradigms such as fog computing will bring an unprecedented technology revolution to our world. 



\section{Real Time Data Utilization Challenges by IoT}

IoT devices are gradually entering into different aspects of human life. Personal health monitors as smart-watches \cite{applewatch,fitbit,samsung}, automated interconnected home appliances, deep-learning driven IoT transformations in medical diagnosis, efficient shopping, defect detection in industries \cite{inteliot} are some of the applications of Iot to name.

According to Yasumoto et al. \cite{2016_yasumoto_survey}, wireless sensor networks [20] yield multiple IoT data streams that are processed by appropriate software applications. For transmission of IoT data streams various network protocols are used for instance Message Queue Telemetry Transport (MQTT) \cite{MQTT}, Constraint Application Protocol (CoAP) \cite{CoAP}, Web Socket \cite{web}, IPv6 over Low Power Wireless Personal Area Networks (IPv6LoWPAN) \cite{6LoWPAN}, MINA \cite{MINA} etc. Such protocols often give rise to edge-heavy computing which needs extension to Fog computing \cite{ref_papluan} to relieve the pressure of processing and storage near the node devices. The next step is data stream processing. On the basis of \cite{FIFO}, the processing is articulated as FIFO queues that respond to data streams and operators i.e. data stream processors handle multiple inputs to and outputs from the FIFO. Different IoT platforms, often supported by machine learning strategies, exist for high-speed real-time processing of vast amounts of temporal data such as IBM Infosphere Stream \cite{stream}, Amazon AWS IoT, Apache Storm[45] and Spark \cite{spark}, Microsoft StreamInsight etc. 

However, by reference of \cite{2016_yasumoto_survey} IoT performance is yet lagging in the fields of heterogeneous data processing, availability of ample IoT platforms with dynamic allocation of resources and granularity customization to mitigate the bandwidth constraint, ability of most platforms to provide automated curation matching and understanding the human version of curation and privacy protection policies applicable for various forms of data. Owing to the problems, the following use-cases cannot be sufficed by current IoT platforms successfully: Firstly, participatory live street view for tourist direction and security services through user, vehicular and street cameras cannot be implemented because cloud resources and network bandwidth are drained out by the incessant upload and downloads of video to the cloud. A new framework allowing parallel flow of multiple data streams to cloud and among stakeholders is necessary instead of convergence. Secondly, ultra-realistic live sports broadcasts through user generated contents is a challenge yet to be resoled due to lack of intelligent and valuable curation of meaningful data from data streams. Thirdly, real-time pedestrian flow tracking in cities for obstruction less vehicular shifting and emergency evacuation is particularly difficult. Not only the latency behind uploading information and processing at the cloud changes the flow of crowd but also the distribution of information to places beyond the location of generation becomes unscalable. Last but not the least, in the monitoring system of the senior citizens, privacy and anomaly detection is of superior concern. Althogh the activity recognition system by Ueda et al. \cite{Ueda} detects 11 activities with 90\% accuracy, the offline learning strategy without the balance of near the edge processing and cloud storage, makes it less effective. Hence 5G should enable the IoT platforms to abstain from the limitations to enhance performance.

\section{Machine Type Communications Structure Redesign}
Conduloci et al. \cite{ref_papcon} emphasized on the machine-type multicast service (MtMS) to reconciliate the end-to-end reliability, latency and energy consumption in the up-downlinks. In case of real time applications for instance, smart city, traffic and pollution sensors collect huge amount of data every second. The highly spatial information is later big data processed which in turn leads to highly reliable and fast decisions like changing traffic lights. For such real-time events, connectivity technologies that provide service level agreements (SLAs), i.e., cellular 3GPP technologies \cite{ref_pappal} such as Long Term Evolution (LTE) \cite{ref_papeutra} and beyond 5G \cite{ref_papboc} systems are necessary. 

The end-to-end delay in MTC environments is caused by uplink(UL), core network and downlink(DL). The UL is utilized to transfer data to remote servers. While the network is in idle state, the data is transmitted through random access (RA) procedure. If two devices send the same preface in the same RA slot, collisions occur. Such collisions and induced delays can be circumvented by access class barring (ACB) \cite{ref_papevo} and extended access barring (EAB) \cite{ref_papran} which introduce backoff mechanism. Although the short access delays to high-priority devices may decline, higher delays incur to other devices. The delays could be reduced through the addition of more preambles rather than only one in the RA as it would abstain from RA reattempts \cite{ref_papthom}.

Management of control and data traffic causes delays in the core network. Although the delay in core network is relatively small, the core network adds delays in the end-to-end communication in the UL and DL directions. Scheduler, frame size, retransmission delay and waiting time for the next transmission frame, round-trip delay in the LTE network act as agents to increment the time by 10-20 ms. The additional delay by the core network in this is as minimal as 1 ms \cite{ref_papbla}. Activation of UE for UL transfer with or without minimal intervention \cite{ref_papenh}, softwarization, and virtualization in the mobile core of 5G system architecture \cite{ref_papli,ref_papwood,ref_papmij,ref_papaiss} have shown promising performance in reducing overhead and delay, but these approaches need further examination.

In the DL direction of mobile communication, paging procedure \cite{ref_papmob} is used to alert the MTC devices before data traffic gets delivered. But the capacity of 3GPP to accommodate devices is insufficient. Only 16 devices can receive the page while receiving data. Aiding the problem, there is provision of only two paging occasion in a radio frame of 10ms. High overhead occurs due to large number of control messages when the number of devices receiving page rises. In order to ameliorate the problem, group paging has been introduced to page a group of devices holding a unique group identity (GID) \cite{ref_papran,ref_papperf}. But the group paging technique has only been tested coupling it with the legacy 3GPP RA procedure, which in fact, cannot process concurrent multiple entries and in turn increases the collision probability. The back-off methodology for ACB/EAB approaches in UL can be modified and used for DL\cite{ref_papevo,ref_papgp1,ref_papgp2}. However, as in one way the back-off values reduce the chances of collision, in the other way, the delay increases.

The concept of group-oriented services led to the Multimedia Broadcast Multicast Services (MBMS) \cite{ref_papevomult} which can endorse group of devices simultaneously for instance mobile TV, video streaming, multimedia content download over mobile networks \cite{ref_papmulti}. The bigger picture is particularly helpful for 5G Architecture to realize the idea of a smart world i.e. smart cities, smart homes, industrial plants, intelligent transportation systems, etc. \cite{ref_papiotind} There are two aspects to be considered to successfully direct the MTC traffic. Firstly, the standard of MBMS is session-oriented. The network provider is responsible for managing a particular MBMS session in precise areas among definite group of devices under a specific client of its own network. Secondly, MBMS is a human-oriented standard. For the formation of multicast groups, human interaction and response is necessary while processing the publication of the MBMS session and joining requests for the session. Additionally, the control traffic has to be redesigned in order to cut the delay of MTC traffic (few bytes) incorporated with the control traffic in the MBMS session.

The design of machine-type multicast service (MtMS) centers the service capability server (SCS) because it handles all the data transmission with MTC device and provides the client with the access to control devices in the group. This decision is particularly advantageous not only in resource utilization in the network but also in reducing delay, overhead and energy consumption in MTC devices. MtMS serving center (MtMS-SC) launches the MtMS session with the list of devices and multicast content at the MtMS gateway (MtMS-GW). The joining request is activated through the the mobility management entity (MME) that provides the MtMS coordination entity (MtMS-CE) with the tracking area information of the devices that will receive page with the help of M3 interface. After completion of the joining request, the MtMS-GW finishes the data transfer through the M1 interface. Finally, the values associated with power, modulation and coding scheme are sent to the involved cells through the M2 interface by the MtMS-CE.

The group-paging and code-expanded RA help the idle state to be aware of MtMS traffic in the joining phase. But the utilization of small channel bandwidth aids to the delay and energy consumption as there is limited resource and time allotment in the radio interface. Enhanced group paging is suggested with subgroups of MtMS group to reap the benefits of lower overhead from the group paging. Furthermore, the time interval between two paging message in an RA frame needs to be fixed carefully to keep the overhead delay minimum in 3GPP. Most importantly, 5G will bridge the small bandwidth problem offering a higher frequency range. So MtMS group-paging will hopefully reap the highest profits from the 5G system architecture. The number of paging requests, total delay for paging, average delay for data delivery and data transmission for standard 3GPP paging (SP), group paging (GP) and enhanced group paging (eGP) are considered as the performance metrics here. According to Conduloci et al. \cite{ref_papcon}, firstly, eGP reduces the number of paging requests by 56\% than SP for 500 devices (32 requests for SP) and further for larger amount of devices. Secondly, in the total delay for paging, the performance of Standard RA (S-RA) and Code expanded RA (CE-RA) are considered. The limiting aspect of S-RA for all is exhibited in providing multicast service to large number of devices. On the other hand, if broader bandwidths are available for instance, under 5G service GP/CE-RA and Egp/CE-RA will face reduced delays. Thirdly, in case of eGP/CE-RA the average delay in delivery does not escalate with the increase in number of devices and hence proposes the lowest delay, overhead, and energy consumption. Lastly, as more resources are required for data delivery at DL, full coverage is ensured in multicast services until resources for delivery is equivalent to a threshold (25 in reference).

There are some open issues in reduction of data transmission delay that require further focus in future. Firstly, the delay caused in joining procedure to connect the mobile core can be eliminated by migrating the device information to the home-evolved NodeBs gateway through softwarization and virtualization \cite{ref_papli,ref_papwood,ref_papmij}. Secondly, an edge-cloud can reduce the end-to-end delay by authorizing the activation of MtMS directly in the edge-cloud. Thirdly, the DL can reduce delay and overhead by analyzing data regarding traffic, network load etc. from the UL. Fourthly, in small channel bandwidth, the appropriate number of UEs to be paged in MtMS should also be figured to reduce the delays. Fifthly, the joining mechanism by a DRX Cycle [17] should be investigated so that all the members of a subgroup in the same DRX Cycle could be paged together. The harmony of UL and DL traffic should be ensured by managing traffic priority. Lastly, parameters like residual battery charge in MTC devices should be taken into account for allocating resources and reduce energy consumption.

\section{5G Communication Networks}
\subsection{Overview}
According to the study of Lifeware and Niu et. al \cite{ref_url4,2015_millimeter_wave_comm}, the 5th generation network utilizes the unique features of the high frequency radio spectrum (in the 30GHz-300GHz range) called millimeter waves. Consequently, in 5G communication data transmission will take place at high speed to an enormous extent. Moreover, existing cellular data sparsely occupy the 5G signals. This will enable the signals to be used on account of increasing bandwidth demands in the future. The assistance of small cell to concentrate the network in a smaller range of area and enhance the performance of 5G transmission \cite{2014_small_cells_mimo}. Furthermore, massive mulitple-input-multiple-output service (mMIMO) will also improve the spectral efficiency through vast increment of antennas and sustain uninterrupted signal reception \cite{2014_small_cells_mimo,2018_5g_enabling_tech_benefits_feasibility_limitations_full_duplex_mMIMO}. Large number of antennas in return will attribute in large beamforming returns. The gain will subsidize the collapse of intercell and inter-stream interference and heighten spectral efficiency.

The 4G towers disperse data in an omnidirectional manner. This phenomenon wastes energy and power in beaming the radio waves at locations where internet access is not required. The 5G towers restrict this occurrence by line-of-sight communications.

5G can accommodate over 1,000 more devices per meter than what was possible by 4G. As 5G signals form shorter wavelengths, the size of antennas has shrunken withholding the precise directional control. Hence, the same base station will now be capable of supporting more antennas and conclusively, way more devices. Moreover, the nearly doubled channel capacity will enable full duplex technology \cite{2015_full_duplex_techniques,2018_5g_enabling_tech_benefits_feasibility_limitations_full_duplex_mMIMO}. It contributes in strengthening 5G and contributing to high spectral efficiency.

The end to end latency will be reduced to almost 1ms in 5G from the around 60ms latency in 4G. According to EDN Network \cite{ref_url5}, the 4G LTE cannot bridge this huge gap of latency for three reasons. Firstly, the minimum size for a radio transport block constitutes a sub frame of 1ms length. This 1 ms duration is entirely spent in the transmission of the block through air interface, without considering processing time at devices and network induced transmission latency. On the contrary, 5G applies a scalable version of orthogonal frequency-division multiplexing with varied numerologies. A 1 ms duration can be accommodated by six different slot configurations of 1, 2, 4, 8, 16 and 32 slots. Hence, if 32 slot configuration is chosen, the minimum size of a radio transport block can be reduced to 0.03125 ms. Secondly, the latency reduction in 4G LTE is obstructed due to the allocation delay of resources between device and base station. The semi-persistent scheduling (SPS) is a unique feature in LTE for periodic data transmissions e.g. voice over IP (VoIP) services. When a device intends to make a data transaction, a resource grant procedure is generated and sent. The time gap between a resource scheduling request and sending data packets is at least 8 ms. However, the request scheduling could be excluded if a mobile handset can utilize the periodic resources while a base station allocates SPS resources. The device can start transferring data through the preconfigured periodic resource after receiving the data. Thirdly, the LTE SPS setup solely supports a single device. In the event of conditional use of periodic resources e. g. while providing a collision warning by the device, the periodic resources dedicated to the device are wasted as they remain unused during the other timelines. In order to alleviate the aforementioned problems, 5G adopts a special routine based on the LTE SPS service, to share the periodic resources among multiple devise known as a configured grant. The configured grant allocation removes the packet transmission delay incurred due to scheduling request and hence the utilization ratio of allocated periodic resources escalates. As the base station allots multiple users the configured grant, the users or devices randomly avail the resources when they are ready for data transmission.

The benefits of using 5G will touch every sphere of life with a vast impact. 5G can access the currently unreachable places like deep underground, remote, far-away locations. Sensors can be placed there and in times of calamity people can be alarmed. Smart cars, and health monitors will be capable to respond from remote locations to any abnormal situation and alert the user to take necessary decisions with minimum latency \cite{ref_papov}. Furthermore, in China, a surgeon performed surgery on an animal by telepresence through 5G network \cite{ref_urluknews}. Most of the concepts that seemed hypothetical and experimental before, will enter into the common world of human lives with the arrival of 5G. 

\subsection{Challenges and Concerns}
\subsubsection{Architectural Compromise of 5G and Restoration} 
In 5G a successful connection between the antenna and the end device, is initiated by a clear, line-of-sight propagation is required which complicates the network architecture \cite{2015_millimeter_wave_comm,ref_url4}. Beam training is required for transmitting data along the direction of sender and receiver. Moreover, as mmWaves have smaller cells, many of them do not cover great distances as they get absorbed by humidity, rain and blocked by mobile humans even, degrading the network performance.
Nonetheless, as 5G uses millimeter waves, the range of data distribution shrinks to almost less than 2\% of the range of 4G \cite{5G_mm_wave}. Hence, mass installation of 5G towers needs to be implemented in order to ensure reliable transmission of data. Among the other factors, small cell induces self interference and in mMIMO, more antennas produce orthogonal pilot overhead. The overhead exhausts the radio resources. Several approaches to alleviate self interference and architectural framework designs for small cells and MIMO have been proposed in \cite{2014_small_cells_mimo,2016_5g_small_cell_nets_fronthauling,2014_MIMO_beamforming,2015_full_duplex_techniques,2018_5g_enabling_tech_benefits_feasibility_limitations_full_duplex_mMIMO} to complement 5G.
\subsubsection{Impact on Health}
5G uses millimeter waves for data transmission and radiofrequency radiation (RF) is generated as a by product of the use of the wireless technology in phones, wearables and computers. We provide a summary of different views and opinions by scientists and environmentalists in order to articulate how exposure to RF can affect human health. Scientists presume that 5G will cause health issues like abnormal cell division and cell destruction. According to the study \cite{5G_IOT_Health}, RF radiation is carcinogenic \cite{2018_carcinogen} and induces tumors \cite{tumor}, cancer \cite{2019_Cancer,2018_Cancer}, disrupts gene expressions \cite{ge}, motor skills, memory and attention \cite{motor}, interacts with sweat glands \cite{duct1,duct2} etc. NYU WIRELESS \cite{nyu_study} propose the first temperature-based utilizing magnetic resonance imaging (MRI)-based systems to measure the thermal change. However, the investigation of Guraliuc et. al and Koyama et. al \cite{60GHz,eye} perceives an appropriate human body model for dosimetric analysis in 60 GHz band and confirms no significant statistical change in the Human corneal epithelial and human lens epithelial cells in prolonged exposure to RF radiation.
\subsubsection{Privacy concerns}
Due to the design of 5G networks, various privacy and security breach issues arise \cite{ref_papov}. Firstly, the access point selection theorems at the physical level of 5G, pose a threat towards user location leakage. Violation of location privacy can bring forth semantic information attacks, timing attacks, and boundary attacks. Secondly, International Mobile Subscriber Identity (IMSI) from user equipment can expose the identity of the subscriber. Tracking of  Thirdly, 5G integrates various actors for instance, virtual mobile network operators (VMNOs), communication service providers (CSPs), and network infrastructure providers. Each actor has different preference for security and the synchronization among them can be daunting. Fourthly, in 5G, dependence on the new actors e.g. CSP, VMNO etc. leads to sharing the same infrastructure among the actors and security compromise. Hence, the 5G operators will lose governance on the data security and user identity. Fifthly, 5G will liberate the physical boundaries from cloud storage. There will be no guarantee of location of data storage. As different countries enact different privacy policies, data privacy might be violated being stored in the cloud of another country.
\subsubsection{Security Countermeasures}
The 5G architecture requires superior authentication and reliability levels in order to protect the users from security breach. Firstly, according to Norman et. al \cite{IMSI}, IMSI can be secured by the establishment of a pseudonym locally from the user equipment and the home network. The scenario also ensures recovery from lock-out in case a stakeholder has lost the pseudonym.

Lastly, the government should enact the privacy policies according to the need of the country with the collaboration from multinational organizations for instance, the United Nations (UN) and European Union (EU) \cite{ahmad_5G}. Additionally, industry as well as consumer-level regulations require to be ensured, so that groups and individuals can design and enjoy appropriate and necessary levels of privacy and security \cite{ahmad_5G}.

\section{On Fog Computing}

As the network of ubiquitous IoT devices grows and as new smart applications are developed, questions beyond scalable communication between devices and connectivity are brought forth. In particular, we must consider that as the number of devices increases, the network is inundated with huge amount of pragmatic data \cite{ref_papdast}.

The data load is burdensome for traditional storage systems and analytic applications. Hence, the introduction of cloud computing served as a solution that offers scalable processing capacity and on demand storage. However, cloud-only IoT architectures have extensive infrastructure and connectivity limitations. In particular, the process of sending and requesting all the data from the cloud, for either storage or compute, is inadequate for emergency situations. For critical highly-responsive applications, low latency and high throughput of data is a requirement, and a high reliance on the cloud represent a risk. In addition, the scalability will be reduced and network bandwidth faces saturation. 

A relative recently proposed solution has been to selectively move computation, storage and control closer to the network edge where data is produced by introducing a system-level architecture. Although, no formal definition has been accepted by the community we can think of the latter as Fog Computing \cite{ref_papedge}. Fog computing offers a balance between cloud and edge computing - accessing the cloud to eliminate resource contention and utilizing the geographically distributed edge devices when needed. It promotes and balances the programming, communication, analytics, and storage among data centers and end devices. To ensure low latency, it accommodates user mobility, heterogeneous resources and interfaces, and distributed data processing.


According to Bonomi et al. \cite{ref_papbonomi} Fog computing is necessary in different fields for instance: 
\begin{itemize}
\item In cloud service, the implementation details are omitted which is often deemed valuable to reduce latency. However, premium latency applications like gaming, live streaming in virtual/augmented reality, video conferencing etc. require the liberty of accessing the accurate information of where processing or storage occurs if they require. 
\item Distributed application in different geographical locations as pipeline auditing, environmental change monitor etc.
\item Mobile appliances with high speeds as smart cars, connected rails etc.
\item Large-scale parallel control systems as smart grid, smart traffic light system etc.
\end{itemize}

Peng et al. \cite{ref_pappeng} propose a fog computing based radio access network(F-RAN) in 5G network that will relieve the pressure on fronthaul and  baseband unit pool and its fast and budget efficient scaling helps F-RAN adjust with the dynamic traffic and radio environment. Tran et al. \cite{ref_paptran} improves the performance of F-RAN in 5G by formulating a framework that homogenizes the heterogeneous resources at the edge collaborating Mobile Edge Computing (MEC) servers and mobile devices. The advantages of incorporating MEC with Fog computing are demonstrated in the fields of mobile-edge orchestration, collaborative caching and processing, and multi-layer interference cancellation. However, edge caching (already solved and improved by MEC), software-defined networking (SDN), network function virtualization (NFV) \cite{ref_pappeng,ref_papluan}, resource management, interoperability, service discovery, mobility support, fairness, security are some of the open-ended fields for 5G in future studies.

Overall, 5G can be seen as an enabler of fog computing, providing ultra-low latency and high bandwidth to communicate and process data anywhere at the edge of the network. A Fog Computing architecture, in turn, aims provides IoT systems with a smarter and efficient way to interact with compute and storage resources. Furthermore, it can be seen as not just suitable due to technology requirements but also due to business requirements that demand regulatory control over resources. As noted by \cite{2016_openfog_architecture}, the use of a fog nodes and cloud resources can depend on the domain specific scenario and application. This benefit can be noted when we consider the application of fog computing to enable data processing, e.g AI, closer to data sources such as remote sensors in a smart city. Fog computing could not only reduce the turnaround time of processing data and reduce network costs, which can be heighten in a cloud-based architecture, but also assist in the management of data and devices within a specific jurisdiction.

Currently, the OpenFog Consortium has developed a reference architecture that has been adopted as an official standard by IEEE \cite{2016_openfog_architecture}. This reference, known as IEEE 1934, defines eight core technical pillars: security, scalability, openness, autonomy, RAS (reliability, availability and serviceability), agility, hierarchy and programmability.

\section{Enterprise IoT Platforms}
Corno et al. \cite{corno} have analyzed the services and performances of the current significant IoT platforms based on eight distinguished features- data storage, devices SDK, mobile SDK, push notifications, Rest APIs, supported protocols, virtual devices, and analytics. The comparison also provides an overview of what services the IoT platforms can deliver and their applications.

Xively, now empowered by google, employs a unique IoT platform as a service \cite{2016_ray_survey}. It provides template application for mobile phones \cite{corno}. The platform does not provide data storage at its end and the push notication for generating alerts is minimum \cite{corno}. Xively uses the MQTT, HTTP protocols and MQTT provides the virtual support for device twin \cite{corno}.

Bosch and Arrayent IoT platforms provide limited functionality \cite{corno}. Bosch lacks the data storage, and Mobile SDK support. On the other hand, Arrayent IoT cloud services do not support virtual devices, hardware SDK and storage of data at the platform. For Bosch IoT suite, HTTP, MQTT, LWM2M, mPRM network protocols are followed; IoT remote manager handles the hardware SDK; Java client or HTTP API Manager integrate with software applications; IoT developer console performs data analysis; push notifications are generated for remote events \cite{corno}. Arrayent abides by HTTPS, and web sockets as protocol; is accessed by mobile phones through iOS and android; its REST API follows EcoAdaptor framework; covers data analytics service; generates realtime alerts \cite{corno}.

AWS IoT Core, Google Cloud IoT, IBM Watson IoT, Microsoft Azure IoT, Oracle IoT Cloud Service, and SAP IoT platforms offers different generic cloud services along with IoT ecosystem \cite{corno}. The push notification is mostly generated by MQTT subscription \cite{corno}. They offer mostly the same set of sevices \cite{corno}. 

Google Cloud IoT platform follows the MQTT, and HTTP protocols \cite{corno}. It accommodates both hardware and mobile SDK with the REST support from Google Cloud IoT API. Cloud Pub/Sub (7 days) implement the device twins \cite{corno}. Google has its own cloud and Data Studio for big data analysis \cite{corno}. Lastly, Cloud Pub/Sub generates push notifications \cite{corno}. 

SAP IoT platform does not follow specific protocols \cite{corno}. It provides hardware support, Mobile SDK is enterprised by cloud platform and iOS extended by REST API and storage at the platform \cite{corno}. It uses the Apple Push Notification services for alerts and messages \cite{corno}. SAP utilizes the Thing Registry for virtual device \cite{corno}.

General Electrics has a more focused set of functions \cite{corno}. It follows the MQTT, HTTP protocol  \cite{corno} for network. For Hardware and Mobile SDK support, there is provision for hybrid Predix Machine \cite{corno}. The REST API is bolstered by the asset services \cite{corno}. Mobile gateway acts as the device twin and data is stored at Blobstore \cite{corno}. However, data analysis and push notification services are missing \cite{corno}.

IBM Watson IoT follows the same MQTT and HTTP as the General Electrics and the Google Cloud IoT platform \cite{corno}. Data analysis is performed by MQTT Watson IoT and data is stored at the Bluemix storage \cite{corno}. Push Notification is developed by Bluemix Push Notifications \cite{corno}. It supports Edge analytics SDK in Hardware respect and Android applications for mobile. Virtual devices are implemented through MQTT \cite{corno}.

The thinger.io IoT platform abides by the HTTPs protocol \cite{corno} and as hardware SDK, can access Arduino, Sigfox or Linux SDK and Android applications alongwith Server API for mobile \cite{corno}. It can access device twin virtually \cite{corno} and contains data bucket \cite{corno}. Data analytics is performed by cloud console \cite{corno} and has no push-notification service \cite{corno}. 

Micosoft Azure IoT pursues AMQP along with the popular MQTT and HTTPS \cite{corno}. The device provisioning strategy is implied at the hardware sector and both Android and iOS applications are supported \cite{corno}. Besides, data analysis, it also stores data at the Azure storage \cite{corno}. It uses IoT edge to develop device twin and alerts the cloud through Notification hub \cite{corno}.

Oracle IoT Cloud Service supports device twinning \cite{corno}, stores data at the platform and allows notification to the cloud \cite{corno}. It follows MQTT and HTTPs as protocols \cite{corno}. It constitutes Endpoint management to handle hardware and uses Java and iOS to access mobile phones \cite{corno}.

AWS IoT core anticipated as one of the pioneering data clouds and IoT platform, is highly conducive for developing industries \cite{aws_azure1,aws_azure2}. It follows HTTP, MQTT and web-socket for network protocols. It can access both android and iOS applications and connects hardware via AWS Greengrass \cite{corno}. The data is stored in S3 and analysed at AWS Console. For any event, notifications are sent to the cloud through Amazon SNS. 

Last but not the least, for fast and reactive smart applications, the reputation of Cisco is rising day by day \cite{cis,smart_city_cis}. It is the only platform that supports Fog computing till date \cite{cis}. It supports independent hardware vendors \cite{smart_city_cis} and can allow access to iOS applications through programmable APIs \cite{cis_archi}. It utilizes gateway to reinforce virtual device \cite{cis_kin}. Besides, data storage and analysis operations, Cisco IoT platform sends notification to cloud on instance of an event \cite{smart_city_cis,cis,cis_archi}.

\section{Implication of 5G on IoT Platforms}
In the following table we investigate and compare the aforementioned IoT Platforms on the basis of their fundamental features and a 5G component- Fog Computing.

\begin{figure}[!ht]
\caption{Relative Study of the Features of Popular IoT Platforms}\label{tab1}
\includegraphics[width=\textwidth]{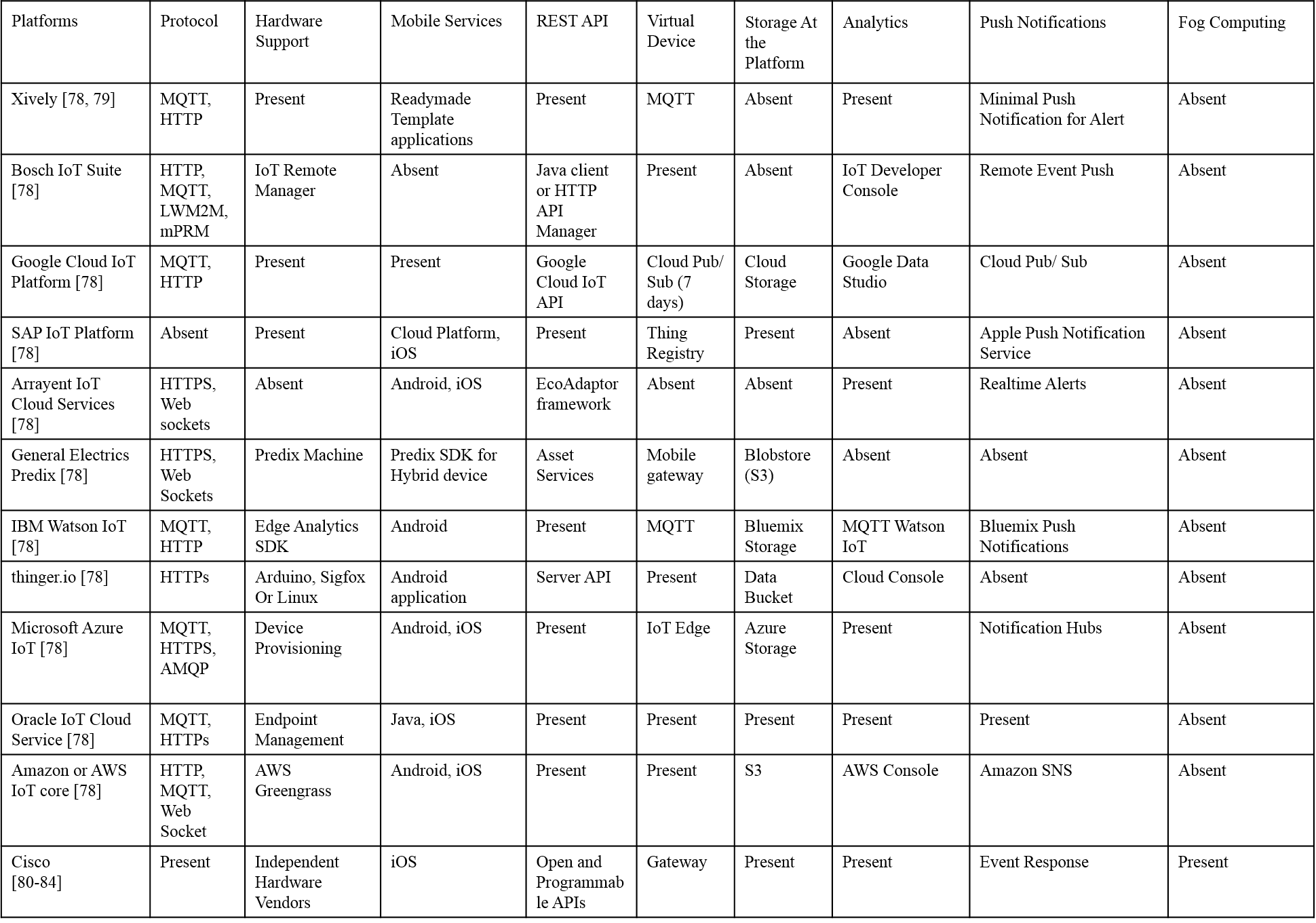}
\end{figure}

From Table:\ref{tab1}, the lack of implementation of Fog computing in all the platforms except Cisco is easily distinguishable. Currently, Microsoft Azure, and AWS IoT are the most popular platforms for cutting edge innovative applications in industries \cite{aws_azure1,aws_azure2}. But in the event of highly responsive systems making decisions within seconds, Fog computing takes the lead and  is clearly the merger of 4G and 5G. Let us consider smart transportation systems. Commercial Jet Planes generate 10 TB of data per 30 minutes \cite{cis}. If the data were stored in cloud, analyzed and necessary data had to be retrieved, the whole process could occupy a time span of minutes to days. When response is time sensitive (less than a second) and mass data is collected at the edge, Fog computing is the best solution to secure and maintain communication effectively. Hence, Cisco IoT platform is largely accepted to implement smart cities and smart vehicular transportation systems due to its sensitive reactivity and conformity with 5G communication \cite{smart_city_cis,cis}.

\section{Discussion}

The Internet of Things will demand a computing architecture that is more decentralized than current computing and data storage models. 

Among different platforms, Cisco is the pioneer in integrating Fog computing into its feature set. Fog computing balances the computational procedures and storage among the edge and cloud. It is an instinctive characteristic of 5G which will diminish the latency and result in fast communication. In sensitive and reactive systems for instance, commercial jet planes \cite{cis} about 10TB data gets transmitted every half an hour. Propagating to and from cloud the cloud disrupts the fast decision-making and increases latency. Hence smart cars and other intelligent vehicles are using Fog computing. The purpose of the survey is to use a suitable IoT platform for a remotely controlled telepresence robots. The robots will access disastrous environment, rescue the victims and react to critical situations. We shall build and test our robots implementing Cisco IoT platform to fulfill our requirements and be responsive to the surroundings being remotely controlled.

When looking into the future trends of compute services, we must look into Amazon Web Services, Google Cloud, Microsoft Azure and Cisco's Cloud Service. Inherently, these 4 tech giants are also leading the pack in IoT services. From a business perspective, the increasing demands for storage and compute from these devices presents a profitable opportunity. Nonetheless, these tech giants must also transform their technology and business models as the Internet of Things begins to demand for a more decentralized infrastructure that provides higher quality of service. As discussed, fog computing is one technology trend that promises to significantly reduce the amount of processing power required from the cloud and shift it to the edge of the network. 

As the leader in cloud computing \cite{amazon_lead,amazon_leader}, as of this writing, Amazon's IoT services offer far greater number of developer tools and cloud computing features than its rivals, and has a dedicated IoT edge computing service called AWS IoT Edge \cite{aws_iot_edge,aws_iot_edge_gg}. With the AWS IoT platform, devices can communicate with application running in AWS instances like Lambda, and use communication protocol like HTTP, MQTT or WebSockets. As part of their IoT Platform, Amazon offers IoT Greengrass \cite{aws_iot_edge_gg,gg}, which extends AWS to edge devices allowing them to act locally on the data they generate, while still using the cloud for management, analytics, and persistent storage. This, in essence, means that devices connected to Greengrass can respond quickly to local events, interact with local resources and connect to the cloud when needed, e.g. for backing up storage, or extended analytics. Other services include AWS IoT Core which enables simple and secure interaction between devices \cite{AWS_IoT_Core}. Additionally, Amazon provides Amazon FreeRTOS - a microcontroller operating system for low-power edge devices that allows for seamless integration with other Amazon IoT services such as IoT Core and Greengrass.

Similar to Amazon, Google has it's own IoT platform - Google Cloud IoT Platform \cite{gc1,gc2,gc3,gc4}. Google Cloud IoT Core serves as the backbone to Google's IoT service offerings, allowing developer to securely connect, manage and process data from millions of globally connected devices. Similar to Amazon's Lambda Functions, Cloud IoT Core run's on Google's serverless infrastructure, providing an infrastructure that scales and responds to real-tie events. Currently, Cloud IoT Core only support MQTT and HTTP. Similar to Greengrass, Google's Cloud IoT Edge service allows edge devices to respond to real-time events and make decisions without continuously communicating to the cloud. The latter two IoT services can be connected to other Google Cloud services such as Cloud Pub/Sub or Firebase.

A key differentiator for Microsoft Azure IoT service's is its Azure IoT Protocol Gateway which adapts incoming and outgoing traffic to comply to a particular devices communication protocol (e.g. AMPQ, MQTT and HTTP).

The target of the analysis of different platforms further empowered by 5G is to implement one of them in building a telepresence robots. Based on sensitive response, power efficiency, data management, and access to Fog computing, Cisco has been proposed to assuage the purpose \cite{cis_white,smart_city_cis,cis_archi,cis_kin,cis}. As mentioned earlier, only Cisco enacted Fog computing into their platform. Fog computing authorizes a balance in the data storage, assembly, analysis and interpretation between the edge and cloud. The intelligent data processing technique offered by Fog computing will support the telepresence robots to react spontaneously and make decisions to handle dangerous situations swiftly. Moreover, the speed and management of huge volumes of data by 5G network will contribute further to realize the goal of a realtime autonomous as well as instinctive robot adept to dealing with emergencies.

\section{Conclusion}

The Internet of Things represents the vision of an interconnected world of devices and humans. 
The development of fifth-generation communication networks brings us a step closer to such vision by enabling faster and more reliable connectivity. In parallel, it further enables other technological advancement that are necessary to bring forth an interconnected world. It further emancipates the Internet of Things to be preached at the broadened spheres of technological worlds. The scrutinizing of various platforms, provides with an insight of their features and capabilities for different goals. Therefore, Cisco aims at fulfilling the motivation of an environment sensitive and fast-reacting telepresence robots for emergencies. The analytical survey shall provide researchers to gather knowledge on distinct platforms and aid in realizing their projects using the appropriate.

%
%


\begin{thebibliography}{8}

\bibitem{ref_url1}
Featured page of Internet of Things Wiki, \url{https://internetofthingswiki.com/top-20-iot-platforms/634/}.
Last accessed 20 April 2019

\bibitem{ref_url2}
OnQ Blog of Qualcomm, \url{http://tiny.cc/wnyj5y}. 
Last accessed 20 April 2019

\bibitem{ref_url3}
Updated List IoT Platforms in 2017 by IoT Analytics, \url{https://iot-analytics.com/iot-platforms-company-list-2017-update/}. Last accessed 21 April 2019

\bibitem{2010_mattern_internet_of_computers_to_iot}
Mattern, F. and Floerkemeier, C.: From the Internet of Computers to the Internet of Things, In From active data management to event-based systems and more, pp. 242-259, Springer, Berlin, Heidelberg. (2010)

\bibitem{1991_weiser_the_computer}
Weiser, M.: The computer for the 21st century. IEEE pervasive computing, 1(1), pp.19-25. (2002)

\bibitem{2015_iot_framework_smart_energy_in_buildings}
Pan, J., Jain, R., Paul, S., Vu, T., Saifullah, A. and Sha, M.: An internet of things framework for smart energy in buildings: designs, prototype, and experiments, IEEE Internet of Things Journal, 2(6), pp.527-537. (2015)

\bibitem{2015_novel_wireless_sensor_network_urban_transport}
Hu, X., Yang, L. and Xiong, W.: A novel wireless sensor network frame for urban transportation, IEEE Internet of Things Journal, 2(6), pp.586-595. (2015)

\bibitem{2016_iot_connected_nav_sys_urban_bus_riders}
Handte, M., Foell, S., Wagner, S., Kortuem, G. and Marrón, P.J.: An internet-of-things enabled connected navigation system for urban bus riders, IEEE internet of things journal, 3(5), pp.735-744. (2016)

\bibitem{2017_iot_considerations_reqs_archs_smart_buildings_energy}
Minoli, D., Sohraby, K. and Occhiogrosso, B.: IoT considerations, requirements, and architectures for smart buildings—Energy optimization and next-generation building management systems, IEEE Internet of Things Journal, 4(1), pp.269-283. (2017)

\bibitem{2015_millimeter_wave_comm}
Niu, Y., Li, Y., Jin, D., Su, L. and Vasilakos, A.V.: A survey of millimeter wave communications (mmWave) for 5G: opportunities and challenges. Wireless Networks, 21(8), pp.2657-2676. (2015)

\bibitem{2018_5G_milli_wave_propagation_models_and_perf_eval}
Sun, S., Rappaport, T.S., Shafi, M., Tang, P., Zhang, J. and Smith, P.J.: Propagation models and performance evaluation for 5G millimeter-wave bands, IEEE Transactions on Vehicular Technology, 67(9), pp.8422-8439. (2018)

\bibitem{2014_small_cells_mimo}
Jungnickel, V., Manolakis, K., Zirwas, W., Panzner, B., Braun, V., Lossow, M., Sternad, M., Apelfrojd, R. and Svensson, T.: The role of small cells, coordinated multipoint, and massive MIMO in 5G, IEEE Communications Magazine, 52(5), pp.44-51. (2014)

\bibitem{2016_5g_small_cell_nets_fronthauling}
Zhang, H., Dong, Y., Cheng, J., Hossain, M.J. and Leung, V.C.: Fronthauling for 5G LTE-U ultra dense cloud small cell networks. IEEE Wireless Communications, 23(6), pp.48-53. (2016)

\bibitem{2014_MIMO_beamforming}
Vook, F.W., Ghosh, A., and Thomas, T.A.: MIMO and beamforming solutions for 5G technology, In 2014 IEEE MTT-S International Microwave Symposium (IMS2014) (pp. 1-4), IEEE.

\bibitem{2015_full_duplex_techniques} 
Zhang, Z., Chai, X., Long, K., Vasilakos, A.V. and Hanzo, L.: Full duplex techniques for 5G networks: self-interference cancellation, protocol design, and relay selection. IEEE Communications Magazine, 53(5), pp.128-137. (2015)

\bibitem{2018_5g_enabling_tech_benefits_feasibility_limitations_full_duplex_mMIMO}
Xia, X., Xu, K., Wang, Y. and Xu, Y.: A 5G-enabling technology: benefits, feasibility, and limitations of in-band full-duplex mMIMO. IEEE Vehicular Technology Magazine, 13(3), pp.81-90. (2018)

\bibitem{applewatch}
Health options in Apple Watch Series 4,
\url{https://www.apple.com/apple-watch-series-4/health/}. Last accessed 27 April 2019

\bibitem{fitbit}
Health features in fitbit, \url{https://www.wired.com/story/when-your-activity-tracker-becomes-a-personal-medical-device/}. Last accessed 27 April 2019

\bibitem{samsung}
Health Monitoring System in Samsung Gear S3 Frontier, \url{https://support.t-mobile.com/docs/DOC-33560}. Last accessed 27 April 2019

\bibitem{inteliot}
Transformation IoT features page, \url{http://tiny.cc/gyou5y}. Last accessed 28 April 2019

\bibitem{2016_yasumoto_survey}
Yasumoto, K., Yamaguchi, H. and Shigeno, H.: Survey of real-time processing technologies of iot data streams, Journal of Information Processing, 24(2), pp.195-202, 2016.

\bibitem{MQTT}
COASIS Standard, MQTT version 3.1.1, available from
\url{http://docs.oasis-open.org/mqtt/mqtt/v3.1.1/os/mqtt-v3.1.1-os.doc}, 2014.

\bibitem{CoAP}
Shelby, Z., Hartke, K. and Bormann, C.:  Request for Comment 7252, The Constrained Application Protocol (CoAP), available from \url{http://tools.ietf.org/rfc/rfc7252.txt}, 2014.

\bibitem{web}
WebSocket, available from \url{https://www.websocket.org/}. Last accesed on 28 April, 2019

\bibitem{6LoWPAN}
Shelby, Z. and Borman, C.: 6LoWPAN: The Wireless Embedded In-
ternet, John Wiley \& Sons (2011)

\bibitem{MINA}
Qin,  Z.,  Denker,  G.,  Giannelli,  C.,  Bellavista,  P. and Venkatasubramanian, N.:  A Software Defined Networking Architecture for the Internet-of-Things, In: 14th Proceedings IEEE Network Operations and Management Symposium (NOMS), pp.1–9, 2014.

\bibitem{FIFO}
Hirzel, M., Soul\'{e}, R., Schneider, S., Gedik, B. and Grimm, R.: A catalog of stream processing optimizations, ACM Computing Surveys, Vol.46, No.4, Article 46, pp.1–34, 2014.

\bibitem{stream}
StreamBase Systems, available from \url{http://www.streambase.com} (2012). Last accessed on 28 April 2019

\bibitem{storm}
Storm project, available from \url{http://storm-project.net/}
(2012). Last Accessed on 28 April 2019

\bibitem{spark}
Documentation of Apache Spark-Lighting-fast cluster computing, available from \url{http://spark.apache.org/}

\bibitem{Ueda}
Ueda, K., Suwa, H., Arakawa, Y. and Yasumoto, K.:  Exploring
Accuracy-Cost Tradeoff in In-Home Living Activity Recognition
based on Power Consumptions and User Positions, In: 14th Proceedings of IEEE International Conference on Ubiquitous Computing and Communications (IUCC 2015), pp.1131–1137.
\bibitem{ref_papcon}
Condoluci, M., Araniti, G., Mahmoodi, T. and Dohler, M.: Enabling the IoT machine age with 5G: Machine-type multicast services for innovative real-time applications, IEEE Access, 4, pp. 5555-5569, , 2016.

\bibitem{ref_pappal}
Palattella, M. R., Dohler, M., Grieco, A., Rizzo, G., Torsner, J., Engel, T., and Ladid, L.: Internet of Things in the 5G Era: Enablers, Architecture, and Business Models, IEEE Journal on Selected Areas in Communications, vol. 34, pp. 510–527, March 2016.

\bibitem{ref_papeutra}
3GPP: Evolved Universal Terrestrial Radio Access (E-UTRA) and
Evolved Universal Terrestrial Radio Access Network (E-UTRAN); Overall description, TS 36.300.

\bibitem{ref_papboc}
Boccardi, F., Heath, R. W., Lozano, A., Marzetta, T. L., and Popovski, P.: Five disruptive technology directions for 5G, IEEE Communications Magazine, vol. 52, pp. 74–80, February 2014.

\bibitem{ref_papevo}
3GPP: Evolved Universal Terrestrial Radio Access (E-UTRA); Radio Resource Control (RRC), TS 36.331.

\bibitem{ref_papran}
3GPP: RAN improvements for machine-type communications, TR
37.868.

\bibitem{ref_papthom}
Thomsen, H., Pratas, N. K., Stefanovic, C., and Popovski, P.: Code-expanded radio access protocol for machine-to-machine communications, Transactions on Emerging Telecommunications Technologies, vol. 24, no. 4, pp. 355–365, 2013.

\bibitem{ref_papbla}
Blaji\'{c}, T., Noguli\'{c}, D., and Dru\^{z}ijani\'{c}, M.: Latency Improvements in 3g Long Term Evolution, in Mipro CTI, svibanj, 2006.

\bibitem{ref_papenh}
3GPP: Study on Enhancements to Machine-Type Communications
(MTC) and Other Mobile Data Applications; Radio Access Network
(RAN) Aspects, TR 37.869.

\bibitem{ref_papli}
Li, Y. and Chen, M.: Software-Defined Network Function Virtualization: A Survey, IEEE Access, vol. 3, pp. 2542–2553, 2015.

\bibitem{ref_papwood}
Wood, T., Ramakrishnan, K. K., Hwang, J., Liu, G., and Zhang, W.: Toward a software-based network: integrating software defined networking and network function virtualization, IEEE Network, vol. 29, pp. 36–41, May 2015.

\bibitem{ref_papmij} 
Mijumbi, R., Serrat, J., Gorricho, J. L., Bouten, N., Turck, F. D. and Boutaba,  R.: Network Function Virtualization: State-of-the-Art and Research Challenges, IEEE Communications Surveys Tutorials, vol. 18, pp. 236–262, Firstquarter 2016.

\bibitem{ref_papaiss}
Aissioui, A., Ksentini, A., Gueroui, A. M., and Taleb, T.: Toward Elastic Distributed SDN/NFV Controller for 5G Mobile Cloud Management Systems, IEEE Access, vol. 3, pp. 2055–2064, 2015.

\bibitem{ref_papmob}
3GPP: Mobile radio interface layer 3 specification, core network protocols; Stage 2, TS 23.108.

\bibitem{ref_papperf}
Wei, C. H., Cheng, R. G., and Tsao, S. L.: Performance Analysis of Group Paging for Machine-Type Communications in LTE Networks, IEEE Transactions on Vehicular Technology, vol. 62, pp. 3371–3382, Sept 2013.

\bibitem{ref_papgp1}
Arouk, O., Ksentini, A., and Taleb, T.: Group paging optimization for machine-type-communications, IEEE International Conference on Communications (ICC), pp. 6500–6505, June 2015.

\bibitem{ref_papgp2}
Arouk, O., Ksentini, A., and Taleb, T.: Group Paging-based Energy Saving for Massive MTC Accesses in LTE and Beyond Networks, IEEE Journal on Selected Areas in Communications, vol. PP, no. 99, pp. 1–1, 2016.

\bibitem{ref_papevomult}
Lecompte, D., and Gabin, F.: Evolved multimedia broadcast/multicast service (eMBMS) in LTE-advanced: overview and Rel-11 enhancements, IEEE Communications Magazine, vol. 50, pp. 68–74, November 2012.

\bibitem{ref_papmulti}
Condoluci, M., Araniti, G., Molinaro, A. and Iera, A.: Multicast Resource Allocation Enhanced by Channel State Feedbacks for Multiple Scalable Video Coding Streams in LTE Networks, IEEE Transactions on Vehicular Technology, vol. PP, no. 99, pp. 1–1, 2015.

\bibitem{ref_papiotind}
Perera, C., Liu, C. H., Jayawardena, S., and Chen, M.: A Survey on Internet of Things From Industrial Market Perspective, IEEE Access, vol. 2, pp. 1660–1679, 2014.

\bibitem{ref_url4}
Difference between 4G and 5G, \url{https://www.lifewire.com/5g-vs-4g-4156322}. 
Last accessed 20 April 2019

\bibitem{ref_url5}
Study of reduced latecy of 5G waves by EDN Network, \url{http://tiny.cc/rkwm5y} last accessed on 17 April 2019

\bibitem{ref_papov}
Ahmad, I., Kumar, T., Liyanage, M., Okwuibe, J., Ylianttila, M. and Gurtov, A.: Overview of 5G security challenges and solutions. IEEE Communications Standards Magazine, 2(1), pp.36-43, 2018.

\bibitem{ref_urluknews}
News on Remote Robotic Surgery in China, \url{http://tiny.cc/h4mt5y} last accessed on 25 April 2019

\bibitem{5G_IOT_Health}
5G And The IOT: Scientific Overview Of Human Health Risks, \url{https://ehtrust.org/key-issues/cell-phoneswireless/5g-networks-iot-scientific-overview-human-health-risks/}. Last Accessed 30 Apr 2019

\bibitem{2018_carcinogen}
Miller, A.B., Morgan, L.L., Udasin, I. and Davis, D.L.: Cancer epidemiology update, following the 2011 IARC evaluation of radiofrequency electromagnetic fields (Monograph 102), Environmental research, 167, pp.673-683. (2018)

\bibitem{2019_Cancer}
Hardell, L. and Carlberg, M.: Comments on the US National Toxicology Program technical reports on toxicology and carcinogenesis study in rats exposed to whole-body radiofrequency radiation at 900 MHz and in mice exposed to whole-body radiofrequency radiation at 1,900 MHz, International journal of oncology, 54(1), pp.111-127. (2019)

\bibitem{2018_Cancer}
Peleg, M., Nativ, O. and Richter, E.D.: Radio frequency radiation-related cancer: assessing causation in the occupational/military setting. Environmental research, 163, pp.123-133. (2018)

\bibitem{tumor}
Lerchl, A., Klose, M., Grote, K., Wilhelm, A.F., Spathmann, O., Fiedler, T., Streckert, J., Hansen, V. and Clemens, M.: Tumor promotion by exposure to radiofrequency electromagnetic fields below exposure limits for humans, Biochemical and biophysical research communications, 459(4), pp.585-590. (2015)

\bibitem{ge}
Di Ciaula, A.: Towards 5G communication systems: Are there health implications?, International journal of hygiene and environmental health, 221(3), pp.367-375. (2018)

\bibitem{motor}
Meo, S.A., Almahmoud, M., Alsultan, Q., Alotaibi, N., Alnajashi, I. and Hajjar, W.M.: Mobile Phone Base Station Tower Settings Adjacent to School Buildings: Impact on Students’ Cognitive Health, American journal of men's health, 13(1), p.1557988318816914. (2019)

\bibitem{duct1}
Betzalel, N., Feldman, Y. and Ishai, P.B.: The modeling of the absorbance of sub-THz radiation by human skin, IEEE Transactions on Terahertz Science and Technology, 7(5), pp.521-528. (2017)

\bibitem{duct2}
Betzalel, N., Ishai, P.B. and Feldman, Y.: The human skin as a sub-THz receiver–Does 5G pose a danger to it or not? Environmental research, 163, pp.208-216. (2018)

\bibitem{nyu_study}
NYU Research mmWave Health Effects,
\url{https://wireless.engineering.nyu.edu/mmwave-health-effects/}. Last accessed on 29 Apr 2019

\bibitem{60GHz}
Guraliuc, A.R., Zhadobov, M., Sauleau, R., Marnat, L. and Dussopt, L.: Millimeter-wave electromagnetic field exposure from mobile terminals, In: 2015 European Conference on Networks and Communications (EuCNC) (pp. 82-85). IEEE. (2015)

\bibitem{eye}
Koyama, S., Narita, E., Shimizu, Y., Suzuki, Y., Shiina, T., Taki, M., Shinohara, N. and Miyakoshi, J.: Effects of long-term exposure to 60 GHz millimeter-wavelength radiation on the genotoxicity and heat shock protein (Hsp) expression of cells derived from human eye, International journal of environmental research and public health, 13(8), p.802. (2016)
\bibitem{5G_mm_wave}
Webpage on Challenges of 5G, \url{http://tiny.cc/48sz5y}. Last accessed on 28 Apr 2019

\bibitem{IMSI}
Norrman, K., N\"{a}slund, M., and Dubrova, E.: Protecting IMSI and User Privacy in 5G Networks, In: 9th Proceedings of EAI International Conference on Mobile Multimedia Communications Institute for Computer Science, Social-Informatics, and Telecommunication Engineering,  pp. 159–66. (2016)

\bibitem{ahmad_5G}
Ahmad, I., Kumar, T., Liyanage, M., Okwuibe, J., Ylianttila, M., Gurtov, A.: 5G security: Analysis of Threats and Solutions, IEEE Conference Standards for Communication and Networking, pp. 193–99, Sept 2017.

\bibitem{ref_papdast}
Dastjerdi, A. V., and Buyya, R.: Fog computing: Helping the Internet of Things realize its potential. Computer, 49(8), 112-116, 2016.

\bibitem{ref_papedge}
Shi, W. and Dustdar, S.: The Promise of Edge Computing, Computer, vol. 49, no. 5, pp. 78–81, 2016.

\bibitem{ref_papbonomi}
Bonomi, F., Rodolfo, M., Natarajan, P., and Zhu, J.: Fog computing: A platform for internet of things and analytics, In Big data and internet of things: A roadmap for smart environments, pp. 169-186, Springer, Cham, 2014.

\bibitem{ref_pappeng}
Peng, M., Yan, S., Zhang, K., and Wang, C.: Fog computing based radio access networks: Issues and challenges, arXiv preprint, arXiv:1506.04233, Jun 13 2015.

\bibitem{ref_paptran}
Tran, T. X., Hajisami, A., Pandey, P., and Pompili, D.: Collaborative mobile edge computing in 5G networks: New paradigms, scenarios, and challenges, arXiv preprint, arXiv:1612.03184, Dec 9 2016.

\bibitem{ref_papluan}
Luan, T. H., Gao, L., Li, Z., Xiang, Y., Wei, G., and Sun, L.: Fog computing: Focusing on mobile users at the edge, arXiv preprint, arXiv:1502.01815, Feb 6 2015.

\bibitem{2016_openfog_architecture}
OpenFog Consortium Architecture Working Group, \url{https://www.openfogconsortium.org/wp-content/uploads/OpenFog-Architecture-Overview-WP-2-2016.pdf}. Last Accessed on 29 Apr 2019


\bibitem{corno}
Corno, F., De Russis, L. and S\'aenz, J. P.: On The Advanced Services That 5G May Provide To IoT Applications, In 2018 IEEE 5G World Forum (5GWF), pp. 528-531, July 2018

\bibitem{2016_ray_survey}
Ray PP: A Survey of IoT Cloud Platforms, Future Computing and Informatics Journal, doi: 10.1016/j.fcij.2017.02.001 (2017)


\bibitem{cis_white}
Cisco Kinetic for Manufacturing Harnessing IoT data to boost productivity, \url{http://tiny.cc/nv3y5y}. Last accessed on 29 Apr 2019

\bibitem{smart_city_cis}
Smart City framework by Cisco, \url{http://tiny.cc/bh3y5y}. Last accessed on 29 Apr 2019

\bibitem{cis_archi}
Architecture of Cisco IoT Platform, \url{http://tiny.cc/2e3y5y}. Last accessed on 29 Apr 2019

\bibitem{cis_kin}
Homepage of Cisco Kinetic, \url{http://tiny.cc/092y5y}. Last accessed on 29 Apr 2019

\bibitem{cis}
Fog Computing and the Internet of Things: Extend the Cloud to Where the Things Are,
\url{http://tiny.cc/nv3y5y}, 2015. Last accessed on 29 Apr 2019

\bibitem{aws_azure1}
Comparison between Microsoft Azure and Amazon Web Services, \url{https://stackify.com/azure-vs-aws-comparison/}. Last accesed on 29 Apr 2019
	
\bibitem{aws_azure2}
Overview of Microsoft Azure and Amazon Web Services, \url{https://azure.microsoft.com/en-us/overview/azure-vs-aws/}. Last accesed on 29 Apr 2019

\bibitem{amazon_lead}
Cusumano, M.A.: Cloud computing and SaaS as new computing platforms. Commun. ACM, 53(4), pp.27-29. (2010)

\bibitem{amazon_leader}
Smith, R.: Computing in the cloud. Research-Technology Management, 52(5), pp.65-68. (2009)

\bibitem{aws_iot_edge}
Amazon Makes Foray Into Edge Computing With AWS Greengrass, \url{http://tiny.cc/ft1y6y}. Last accesed on 10 May 2019

\bibitem{aws_iot_edge_gg}
AWS IoT Greengrass, \url{https://aws.amazon.com/greengrass/}. Last accesed on 11 May 2019

\bibitem{gg}
What Is AWS IoT Greengrass?, \url{https://docs.aws.amazon.com/greengrass/latest/developerguide/what-is-gg.html}. Last accessed on 11 May 2019

\bibitem{AWS_IoT_Core}
AWS IoT Core Documentation, \url{https://docs.aws.amazon.com/iot/index.html}. Last accessed on 11 May 2019

\bibitem{gc1}
CLOUD IOT CORE, \url{http://tiny.cc/3e3y6y}. Last accessed on 15 May 2019

\bibitem{gc2}
Google Cloud IoT, \url{https://cloud.google.com/solutions/iot/}. Last accessed on 15 May 2019

\bibitem{gc3}
IoT in the Google Cloud, \url{https://www.qwiklabs.com/quests/49}. Last accessed on 15 May 2019

\bibitem{gc4}
Cloud IoT Core overview, \url{https://cloud.google.com/iot/docs/concepts/overview}. Last accessed on 15 May 2019
\end{thebibliography}

\end{document}